\RequirePackage{amsmath}

\documentclass[preprint,12pt]{elsarticle}

\usepackage{bm}
\usepackage{graphicx}
\usepackage{color}
\usepackage{amssymb}
\usepackage{bbold}

\journal{Physica E}

\begin{document}

\begin{frontmatter}

\title{Quantum-confined Stark effect in band-inverted junctions}

\author{A. D\'{\i}az-Fern\'{a}ndez, F. Dom\'inguez-Adame}

\address{GISC, Departamento de F\'{\i}sica de Materiales, Universidad
Complutense, E-28040 Madrid, Spain}

\begin{abstract}

 Topological phases of matter are often characterized by interface states, which were already known to occur at the boundary of a band-inverted junction in semiconductor heterostructures. In IV-VI compounds such interface states are properly described by a two-band model, predicting the appearance of a Dirac cone in single junctions. We study the quantum-confined Stark effect of interface states due to an electric field perpendicular to a band-inverted junction. We find a closed expression to obtain the interface dispersion relation at any field strength and show that the Dirac cone widens under an applied bias. Thus, the Fermi velocity can be substantially lowered even at moderate fields, paving the way for tunable band-engineered devices based on band-inverted junctions. 

\end{abstract}


\end{frontmatter}

\section{Introduction}   \label{sec:intro}

The advent of topology in condensed matter physics has drawn renewed attention to band-inverted semiconductors. These systems were first reported by Dimmock \emph{et al.} in 1966~\cite{Dimmock66}. They showed that the fundamental gap between the bands with symmetries $L_{6}^{-}$ (conduction band) and $L_{6}^{+}$ (valence band) in Pb$_{1-x}$Sn$_{x}$Te decreases monotonically upon increasing the Sn fraction and then reopens with the order of the bands inverted relative to those of PbTe. Nowadays, ternary compounds Pb$_{1-x}$Sn$_{x}$Te and Pb$_{1-x}$Sn$_{x}$Se are known to be topological crystalline insulators~\cite{Fu11,Koumoulis15,Assaf16}. 

Heterojunctions between semiconductors with mutually inverted bands support interface states lying within the gap, provided that the two gaps overlap (see Refs.~\cite{Volkov85,Korenman87,Agassi88,Pankratov90,Kolesnikov97} and references therein). These interface states are protected by symmetry and are responsible for the conducting properties of the surface. From a theoretical perspective, interface states in IV-VI heterojunctions are well described by a two-band model using the effective ${\bm k}\cdot{\bm p}$ approximation~\cite{Kriechbaum86}. The equation governing the conduction- and valence-band envelope functions reduces to a Dirac-like equation after neglecting far-band corrections. In view of the analogy with relativistic quantum mechanics, exact solutions are readily obtained using supersymmetric~\cite{Pankratov90} or Green's function approaches~\cite{Adame94}. A salient feature of interface states is that the interface dispersion is a Dirac cone of the form $E({\bm k}_{\bot})=\hbar v_{F}|{\bm k}_{\bot}|$, ${\bm k}_\bot$ being the interface wave vector. Typically, Fermi's velocity $v_F$ is of the order of $10^{-3}c$ in IV-VI compounds, where $c$ is the speed of light in vacuum. The precise value of $v_F$ depends on the effective mass and the magnitude of the gap. In a IV-VI heterojunction both quantities vary along the growth direction but $v_F$ remains essentially constant~\cite{Agassi88}. 

Device applications demand systems for engineering $v_F$ and traditionally graphene has been viewed as an ideal candidate to achieve this goal~\cite{Hwang12}. Reduction of $v_F$ has been predicted and observed in few-layer graphene due to the rotation of two neighboring layers~\cite{Li10,Trambly10}. Graphene/chlorophyll-a nanohybrids have been put forward as a way towards tuning $v_F$~\cite{Das15}. This hybrid system shows increased electron density and reduced $v_F$ due the appearance of a Van Hove singularity. Moreover, many-body effects can also alter Fermi's velocity. In this regard, a renormalization of $v_F$ in suspended graphene has been related to many-body effects~\cite{Elias11}. This renormalization has also been detected in a topological insulator, namely, Bi$_2$Te$_3$~\cite{Miao13}. Unfortunately, all these mechanisms cannot be dynamically altered in an experiment. In other words, once the sample is grown, there is no way to tune $v_F$ without modifying the structure.

Recently, we have studied band-inverted junctions based on IV-VI compounds using a two-band model when an electric field is applied along the growth direction~\cite{Diaz-Fernandez17}. Assuming symmetric and same-sized gaps, we have demonstrated that the Dirac cone arising in the junction is robust against moderate values of the electric field but becomes wider on increasing the bias. Fermi's velocity was found to decrease quadratically with the applied field. This reduction allows Fermi's velocity to be tuned dynamically and continuously in a controllable way in the same sample. The aim of this paper is to theoretically address the quantum-confined Stark effect in arbitrary-sized but abrupt band-inverted junctions under an electric field of any strength. Results are compared to the analytical predictions of Ref.~\cite{Diaz-Fernandez17} that are only valid for moderate fields. 

\section{Interface states in a band-inverted junction}   \label{sec:edge_states}

Our analysis is based on the effective-mass approximation, which is a reliable method to obtain the electron states near the band edges of IV-VI semiconductors~\cite{Kriechbaum86}. The electron wave function is written as a sum of products of band-edge Bloch functions with slowly varying envelope functions. Keeping only the two nearby $L$  bands, there are four envelope functions (including spin) that can be arranged as a four-component vector ${\bm\chi}({\bm r})$. This vector is composed by the two-component spinors ${\bm\chi}_{+}({\bm r})$ and ${\bm\chi}_{-}({\bm r})$ belonging to the $L^{+}$ and $L^{-}$ bands and subject to an effective Hamiltonian of Dirac form~\cite{Korenman87,Agassi88,Pankratov90} 

\begin{equation}
\mathcal{H}_{0}=v_{\bot}{\bm\alpha}_{\bot}\cdot{\bm p}_{\bot}+v_z\alpha_z p_z
+\frac{1}{2}\,E_{\mathrm{G}}(z)\beta+V_{\mathrm C}(z) \ ,
\label{eq:01}
\end{equation}
where the $Z$ axis is parallel to the growth direction $[111]$. It is understood that the subscript $\bot$ of a vector indicates the nullification of its $z$-component. Here $E_{\mathrm{G}}(z)$ stands for the position-dependent gap and $V_{\mathrm C}(z)$ gives the position of the gap center. ${\bm\alpha}=(\alpha_x,\alpha_y,\alpha_z)$ and $\beta$ denote the usual $4\times 4$ Dirac matrices
\begin{equation}
\alpha_i=\begin{pmatrix}
\mathbb{0}_2 & \sigma_i \\
\sigma_i & \mathbb{0}_2
\end{pmatrix}\ ,
\qquad
\beta=\begin{pmatrix}
\mathbb{1}_2 & \mathbb{0}_2 \\
\mathbb{0}_2 & -\mathbb{1}_2
\end{pmatrix}\ ,
\end{equation}
$\sigma_i$ being the Pauli matrices, and $\mathbb{1}_n$ and $\mathbb{0}_n$ are the $n\times n$ identity and null matrices, respectively. Here $v_{\bot}$ and $v_z$ are interband matrix elements having dimensions of velocity. We take abrupt profiles for both the magnitude of the gap and the gap centre as follows
\begin{eqnarray}
E_{\mathrm{G}}(z)&=&2\Delta_{\mathrm{L}}\Theta(-z)
+2\Delta_{\mathrm{R}}\Theta(z)\ ,\nonumber\\
V_{\mathrm{C}}(z)&=&V_{\mathrm{L}}\Theta(-z)+V_{\mathrm{R}}\Theta(z)\ ,
\label{eq:02}
\end{eqnarray}
where $\Theta(z)$ is the Heaviside step function. The subscripts L and R refer to the left and right sides of the junction, respectively. Note that in the case of a band-inverted junction $\Delta_{\mathrm{L}}\Delta_{\mathrm{R}}<0$.

The interface momentum is conserved and we seek solutions of the form ${\bm\chi}({\bm r})={\bm\Psi}(z)\exp(i{\bm r}_{\bot}\cdot{\bm k}_{\bot})$. The envelope function decays exponentially with distance at each side as ${\bm\Psi}(z)\sim \exp\Big[-K_\mathrm{L,R}({\bm k}_{\bot})|z|\Big]$, where~\cite{Adame94}
\begin{equation}
K_\mathrm{L,R}({\bm k}_{\bot})=\frac{1}{\hbar v_z}\,\sqrt{\Delta_\mathrm{L,R}^{2}-\big[E_{\pm}^{0}({\bm k}_{\bot})-V_\mathrm{L,R}\big]^2
+\hbar^2v_{\bot}^2k_{\bot}^2}\ ,
\label{eq:03}
\end{equation}
and the interface dispersion relation is a Dirac cone
\begin{subequations}
\begin{equation}
E_{\pm}^{0}({\bm k}_{\bot})=V_0 \pm \hbar v_F|{\bm k}_{\bot}| \ ,
\label{eq:04a}
\end{equation}
as long as the gaps overlap, i.e., $(\Delta_{\mathrm{R}}-\Delta_{\mathrm{L}})^2 > (V_{\mathrm{R}}-V_{\mathrm{L}})^2$. The superscript $0$ refers to the field-free junction. The Dirac point is at $V_0$, 
\begin{equation}
V_0=\frac{\Delta_{\mathrm{R}}V_{\mathrm{L}}-\Delta_{\mathrm{L}}V_{\mathrm{R}}}{\Delta_{\mathrm{R}}-\Delta_{\mathrm{L}}}  \ ,
\end{equation}
and Fermi's velocity is given by 
\begin{equation}
v_F =\sqrt{1-\left(\frac{V_{\mathrm{R}}-V_{\mathrm{L}}}{\Delta_{\mathrm{R}}-\Delta_{\mathrm{L}}}\right)^2}\,v_{\bot} \ .
\label{eq:04b}
\end{equation}
\label{eq:04}
\end{subequations}
%

\section{Band-inverted junction under bias}   \label{sec:single}

We now turn to the interface states in a band-inverted junction subject to an electric field ${\bm F}$ applied along the growth direction. The envelope functions satisfy a Dirac-like equation $(\mathcal{H}_0-eFz-E){\bm\chi}({\bm r})=0$. The interface momentum is conserved so that ${\bm\chi}({\bm r})={\bm\Psi}(z)\exp(i{\bm r}_{\bot}\cdot{\bm k}_{\bot})$ still applies. In order to make the presentation of results clearer, we parameterize the gap and gap-center profiles~(\ref{eq:02}) as $E_\mathrm{G}(z)/2=\Delta(z)=\Delta+\lambda \,\mathrm{sgn}(z)$ and $V_\mathrm{C}(z)=V_0+\gamma\Delta(z)$, where $\Delta=(\Delta_{\mathrm{R}}+\Delta_{\mathrm{L}})/2$, $\lambda=(\Delta_{\mathrm{R}}-\Delta_{\mathrm{L}})/2$ and $\gamma=(V_{\mathrm{R}}-V_{\mathrm{L}})/2\lambda$. Let us introduce the length scale of the problem, $d=\hbar v_z/\lambda$, as well as the following dimensionless parameters
\begin{equation} 
\xi=\frac{z}{d} \ , 
\quad 
\bm{\kappa}=\frac{v_{\bot}}{v_z}\,d\,\mathbf{k}_{\bot} \ , 
\quad
\delta = \frac{\Delta}{\lambda}
\quad 
\epsilon=\frac{E-V_0}{\lambda} \ , 
\quad f=\frac{F}{F_{\mathrm{C}}} \ ,
\label{eq:05}
\end{equation}
where $F_{\mathrm{C}}=\lambda/ed$.
Then, Dirac's equation can be written as
\begin{equation}
\Big\{-i\alpha_{z}\partial_{\xi}+\bm{\alpha}_{\bot}\cdot\bm{\kappa}+\beta\delta+\big[\beta+\gamma\big]\mathrm{sgn}(\xi)-\epsilon+\gamma\delta-f\xi\Big\}\mathbf{\Psi}(\xi)=0 \ .
\label{eq:06}
\end{equation}
with $\partial_{\xi}=\text{d}/\text{d}\xi$. We proceed by assuming that the junction is embedded in a very large box of length $2L$. Imposing the current density to vanish at the edges of the box we get $i\beta\alpha_z\mathbf{\Psi}(-\ell)=\mathbf{\Psi}(-\ell)$ and $-i\beta\alpha_z\mathbf{\Psi}(\ell)=\mathbf{\Psi}(\ell)$~\cite{Alberto11}, where $\ell=L/d \gg 1$. Moreover, continuity at the interface amounts to $\mathbf{\Psi}(0^{-})=\mathbf{\Psi}(0^{+})$.

We perform a unitary transformation $\mathbf{\Psi}(\xi)=\mathcal{U}\mathbf{\Phi}(\xi)$ with $\mathcal{U}=(1/\sqrt{2})(\sigma_x+\sigma_z)\otimes \mathbb{1}_2$ that transforms Dirac's equation~(\ref{eq:06}) into
\begin{equation}
\Big\{\sigma_z\otimes\mathcal{H}+\gamma \,\mathrm{sgn}(\xi)-\epsilon+\gamma\delta-f\xi+\big[\delta+\,\mathrm{sgn}(\xi)\big]\sigma_x\otimes\mathbb{1}_2\Big\}\mathbf{\Phi}=0 \ ,
 \label{eq:07}
\end{equation}
where $\mathcal{H}=-i\partial_{\xi}\sigma_z+k_x\sigma_x+k_y\sigma_y$ is nothing but a Dirac Hamiltonian for massless particles. In order to tackle the problem it is convenient to write $\mathbf{\Phi}=(\mathbf{\Phi}_u,\mathbf{\Phi}_l)^{T}$. Doing so, a pair of coupled equations are obtained, which are easily decoupled, resulting in the following equation for the upper component $\mathbf{\Phi}_u$
\begin{subequations}
\begin{equation}
\Big\{\partial_{\xi}^2-\kappa^2-if\sigma_z+\big[\gamma\,\mathrm{sgn}(\xi)-\epsilon+\gamma\delta-f\xi\big]^2-\big[\delta+\,\mathrm{sgn}(\xi)\big]^2\Big\}\mathbf{\Phi}_u=0 \ ,
 \label{eq:08a}
\end{equation}
with $\xi\neq 0$ and $\kappa=|{\bm\kappa}|$. $\mathbf{\Phi}_l$ is then obtained from
\begin{equation}
\mathbf{\Phi}_l=-\frac{1}{\delta+\,\mathrm{sgn}(\xi)}\bigg[\mathcal{H}+\gamma \,\mathrm{sgn}(\xi)-\epsilon+\gamma\delta-f\xi\bigg]\mathbf{\Phi}_u \ .
 \label{eq:08b}
\end{equation}
\label{eq:08}
\end{subequations}
Notice that Eq.~(\ref{eq:08a}) is now diagonal and straightforwardly solved. In fact, one may solve for the upper component of $\mathbf{\Phi}_u$ and obtain the lower component by taking the complex conjugate of the former and different constants of integration. Let
\begin{equation}
 x=\frac{1}{\sqrt{f}}\big[\epsilon-\gamma\delta+f\xi-\gamma \,\mathrm{sgn}(\xi)\big] \ , \quad \mu^2=\frac{1}{4f}\big\{\kappa^2+\big[\delta+\,\mathrm{sgn}(\xi)\big]^2\big\} \ .
 \label{eq:09}
\end{equation}
Then, it can be immediately shown that
\begin{subequations}
\begin{equation}
 \mathbf{\Phi}_u=\begin{pmatrix} M & \sigma_x M^{*} \end{pmatrix}\mathbf{C} \ ,
 \label{eq:10a}
\end{equation}
where $\mathbf{C}$ is a four-component constant vector and $M(x)$ is given by
\begin{equation}
 M(x)=\begin{pmatrix} F^{*}(x) & G(x) \\ 0 & 0 \end{pmatrix} \ ,
 \label{eq:10b}
\end{equation}
and~\cite{Sauter31}
\begin{align}
F(x)&=M\left(-i\mu^2,\frac{1}{2},ix^2\right)e^{-ix^2/2} \ , 
\nonumber \\
G(x)&=-2i\mu x M\left(1-i\mu^2,\frac{3}{2},ix^2\right)e^{-ix^2/2} \ ,
\label{eq:10c}
\end{align}
with $M(a,b,z)$ the Kummer's functions~\cite{Abramowitz72}.
\end{subequations}
The functions $F(x)$ and $G(x)$ satisfy the following useful relations
\begin{align}
\left(i\partial_{x}+x\right)F^{*}(x) & =2\mu G^{*}(x) \ , 
\nonumber \\
\left(i\partial_{x}+x\right)G(x) & =2\mu F(x)\ .
\label{eq:11}
\end{align}
Using these relations and equations (\ref{eq:08b}) and (\ref{eq:10a}) we obtain
\begin{subequations}
\begin{equation}
 \mathbf{\Phi}_l=\begin{pmatrix} \tau M^{*}\sigma_x+\eta\sigma_x M & \tau\sigma_x M\sigma_x+\eta^{*}M^{*} \end{pmatrix}\mathbf{C} \ ,
 \label{eq:12a}
\end{equation}
where we have introduced
\begin{equation}
 \tau=\frac{2\mu\sqrt{f}}{\delta+\mathrm{sgn}(\xi)} \ , \quad \eta=-\frac{\kappa_x+i\kappa_y}{\delta+\mathrm{sgn}(\xi)} \ .
 \label{eq:12b}
\end{equation}
\label{eq:12}
\end{subequations}
Finally, $\mathbf{\Phi}$ can be finally expressed as
\begin{equation}
 \mathbf{\Phi}(x)=\mathcal{F}(x)\mathbf{C} \ , \quad \mathcal{F}(x)=\begin{pmatrix} M & \sigma_x M^{*} \\  \tau M^{*}\sigma_x+\eta\sigma_x M & \tau\sigma_x M\sigma_x+\eta^{*}M^{*} \end{pmatrix} \ .
 \label{eq:13}
\end{equation}

Once the general solution at each side of the junction is known, boundary conditions at the interface and $\xi=\pm\ell$ lead to
\begin{equation}
\det  
\begin{pmatrix}
  \mathcal{F}(x_0^{+}) & -\mathcal{F}(x_0^{-}) \\
  \mathcal{P}^{+}(x_{\ell}^{+}) & (\sigma_x\otimes\mathbb{1}_2)\mathcal{P}^{-}(x_{\ell}^{-})
 \end{pmatrix} = 0 \ ,
\label{eq:15}
\end{equation}
with $x_0^{\pm}=x(\xi=0^{\pm})$, $x_{\ell}^{\pm}=x(\xi=\pm l)$ and
\begin{equation}
\mathcal{P}^{\pm}(x) = \begin{pmatrix} (\eta\sigma_x\pm i\sigma_z)M+\tau M^{*}\sigma_x) & (\eta^{*}\mp \sigma_y)M^{*}+\tau\sigma_xM\sigma_x \\ \mathbb{0}_2 & \mathbb{0}_2 \end{pmatrix} \ .
\label{eq:16}
\end{equation}
%

\section{Numerical results}   \label{sec:results}

\begin{figure}[htb]
\centerline{\includegraphics[width=0.95\columnwidth]{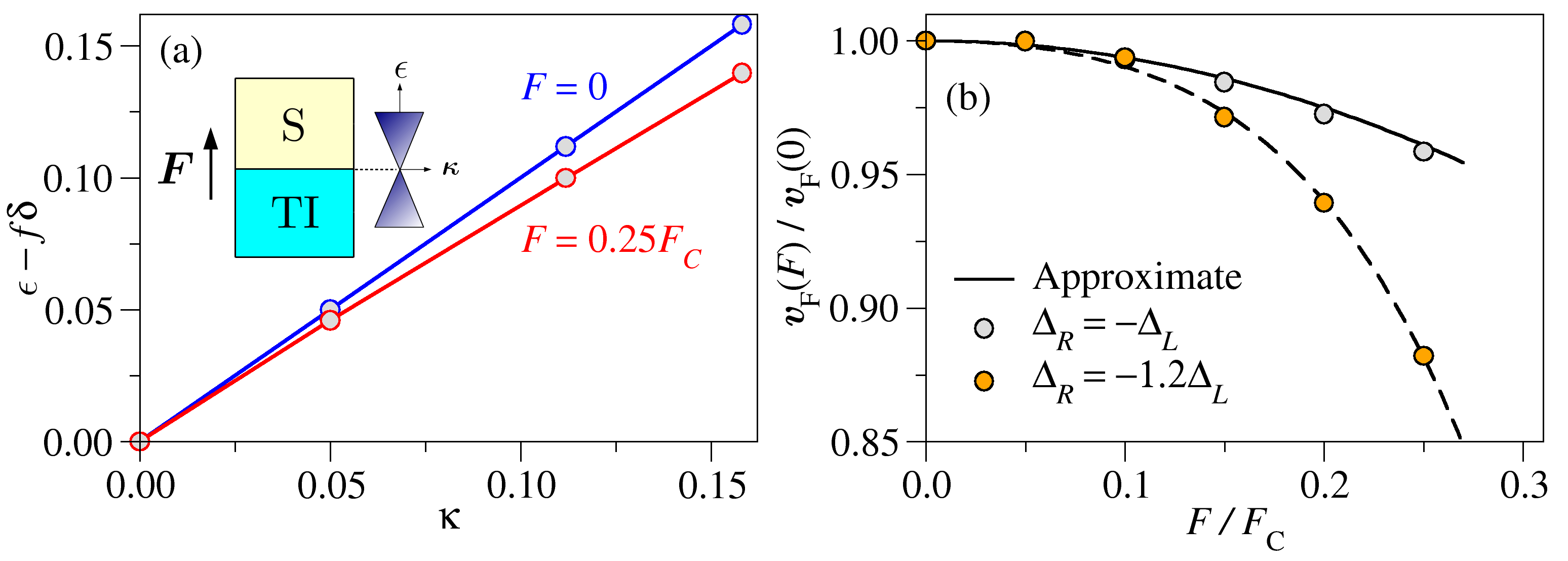}}
\caption{(a)~Energy as a function of the interface momentum for two values of the electric field in an asymmetric junction ($\Delta_R=-1.2\Delta_L$). (b)~Fermi's velocity as a function of the the electric field in an heterojunction with symmetric ($\Delta_R=-\Delta_L$) and asymmetric ($\Delta_R=-1.2\Delta_L$) gap configurations. Solid line corresponds to the approximate result given by~(\ref{eq:18}) for the symmetric configuration. Dashed line is a non-linear fit in powers of $F^{2}$ up to $F^{4}$ for the asymmetric configuration.}
\label{fig1}
\end{figure}

To avoid the profusion of free parameters, in this section we restrict ourselves to band-inverted junctions with centered gaps ($\gamma=0$). Let us start out by calculating the first-order perturbation correction to the field-free dispersion relation~(\ref{eq:04a}). Straightforward algebra yields that to first order in $|\bm{\kappa}|$ and considering $|\delta|\ll 1$
\begin{equation}
\epsilon(\bm{\kappa})=\epsilon_{\pm}^0(\bm{\kappa})+f\delta \ ,
\label{eq:17}
\end{equation}
where $\epsilon_{\pm}^0(\bm{\kappa})=\pm|\bm{\kappa}|$ is the field-free energy, which reduces to equation~(\ref{eq:04a}) when restoring to the original parameters. Therefore, the first-order perturbation approach predicts that the Dirac point shifts upwards or downwards with $f$, depending on the sign of the parameter $\delta$, but Fermi's velocity remaining unaltered. However, that is not the case when numerically solving~(\ref{eq:16}). We found that better numerical accuracy is attained by setting a field-dependent origin of energy, namely after replacing $\epsilon(\bm{\kappa})$ by $\epsilon(\bm{\kappa}) - f\delta$ in~(\ref{eq:16}). While the energy shift of the Dirac point is correctly accounted for by perturbation theory, i.e. $\epsilon(0)-f\delta=0$, the numerical solution of Eq.~(\ref{eq:16}) reveals that the Dirac cone persists but its slope (Fermi's velocity) is lowered at finite values of the reduced electric field $f$. Lowering of Fermi's velocity is clearly seen in Fig.~\ref{fig1}(a), where we compare the interface dispersion relation at $f=F/F_\mathrm{C}=0.25$ with the unbiased junction when the difference in the gap sizes is $20\%$ ($\Delta_R=-1.2\Delta_L$).

In the case of symmetric gaps ($\Delta_R=-\Delta_L$), we were able to obtain an approximate dependence of Fermi's velocity on the electric field, given as
\begin{equation}
v_F(F)=v_F(0)\left(1-\frac{5}{8}\,\frac{F^{2}}{F_\mathrm{C}^{2}}\right) \ ,
\label{eq:18}
\end{equation}
and found that it fits the numerically exact results with outstanding precision even at moderate fields $F \lesssim 0.4F_\mathrm{C}$~\cite{Diaz-Fernandez17}. Figure~\ref{fig1}(b) compares the approximate dependence of Fermi's velocity on the electric field from~(\ref{eq:18}) with the numerical result from~(\ref{eq:16}) when $\Delta_R=-\Delta_L$, confirming the correctness of the former. In the general case of an asymmetric gap we have been unable to arrive at a closed expression similar to~(\ref{eq:18}). Figure~\ref{fig1}(b) also shows the dependence of Fermi's velocity on the electric field when $\Delta_R=-1.2\Delta_L$. We can clearly see a stronger reduction of Fermi's velocity compared to the symmetric gap configuration. In fact, even at moderate fields, the dependence is not quadratic on $F$ but of the form $F^{4}$ (see dashed line).

\section{Conclusions}   \label{sec:conclusion}

In conclusion, we have studied band-inverted junctions under a perpendicular electric field. We used a spinful two-band model that is equivalent to the Dirac model for relativistic electrons. The mass term is half the bandgap and changes its sign across the junction. In view of the analogy with relativistic electrons, we have solved exactly the corresponding Dirac equation that describes the confined Stark effect of the interface states. It is a remarkable result that the interface linear dispersion is preserved and the Fermi velocity is lowered by the electric field. The symmetric gap configuration $\Delta_R=-\Delta_L$ was already discussed in our previous work~\cite{Diaz-Fernandez17}, where it was demonstrated the the lowering of Fermi's velocity is quadratic in the electric field. Remarkably, in this work we found a more dramatic decrease of Fermi's velocity in the general case of asymmetric gaps ($\Delta_R\neq -\Delta_L$). In the range of electric fields discussed in this work, Fermi's velocity decreases as the quartic power of the field and the effect is magnified. The reduction of Fermi's velocity is an effect with measurable consequences on several physical magnitudes, and we expect it to have applications for the design of novel devices based on topological materials. 


\medskip

The authors thank L. Chico for helpful discussions. This work was supported by the Spanish MINECO under grants MAT2013-46308 and MAT2016-75955.

\section*{References}

\bibliography{references}

\begin{thebibliography}{21}
\expandafter\ifx\csname natexlab\endcsname\relax\def\natexlab#1{#1}\fi
\providecommand{\url}[1]{\texttt{#1}}
\providecommand{\href}[2]{#2}
\providecommand{\path}[1]{#1}
\providecommand{\DOIprefix}{doi:}
\providecommand{\ArXivprefix}{arXiv:}
\providecommand{\URLprefix}{URL: }
\providecommand{\Pubmedprefix}{pmid:}
\providecommand{\doi}[1]{\href{http://dx.doi.org/#1}{\path{#1}}}
\providecommand{\Pubmed}[1]{\href{pmid:#1}{\path{#1}}}
\providecommand{\bibinfo}[2]{#2}
\ifx\xfnm\relax \def\xfnm[#1]{\unskip,\space#1}\fi
\bibitem[{Dimmock et~al.(1966)Dimmock, Melngailis, and Strauss}]{Dimmock66}
\bibinfo{author}{J.~O. Dimmock}, \bibinfo{author}{I.~Melngailis},
  \bibinfo{author}{A.~J. Strauss}, \bibinfo{journal}{Phys. Rev. Lett.}
  \bibinfo{volume}{16} (\bibinfo{year}{1966}) \bibinfo{pages}{1193}.
\bibitem[{Fu(2011)}]{Fu11}
\bibinfo{author}{L.~Fu}, \bibinfo{journal}{Phys. Rev. Lett.}
  \bibinfo{volume}{106} (\bibinfo{year}{2011}) \bibinfo{pages}{106802}.
\bibitem[{Koumoulis et~al.(2015)Koumoulis, Chasapis, Leung, Taylor, Stoumpos,
  Calta, Kanatzidis, and Bouchard}]{Koumoulis15}
\bibinfo{author}{D.~Koumoulis}, \bibinfo{author}{T.~C. Chasapis},
  \bibinfo{author}{B.~Leung}, \bibinfo{author}{R.~E. Taylor},
  \bibinfo{author}{C.~C. Stoumpos}, \bibinfo{author}{N.~P. Calta},
  \bibinfo{author}{M.~G. Kanatzidis}, \bibinfo{author}{L.-S. Bouchard},
  \bibinfo{journal}{Adv. Electron. Mater.} \bibinfo{volume}{1}
  (\bibinfo{year}{2015}) \bibinfo{pages}{1500117}.
\bibitem[{Assaf et~al.(2016)Assaf, Phuphachong, Volobuev, Inhofer, Bauer,
  Springholz, de~Vaulchier, and Guldner}]{Assaf16}
\bibinfo{author}{B.~A. Assaf}, \bibinfo{author}{T.~Phuphachong},
  \bibinfo{author}{V.~V. Volobuev}, \bibinfo{author}{A.~Inhofer},
  \bibinfo{author}{G.~Bauer}, \bibinfo{author}{G.~Springholz},
  \bibinfo{author}{L.~A. de~Vaulchier}, \bibinfo{author}{Y.~Guldner},
  \bibinfo{journal}{Sci. Rep.} \bibinfo{volume}{6} (\bibinfo{year}{2016})
  \bibinfo{pages}{20323}.
\bibitem[{Volkov and Pankratov(1985)}]{Volkov85}
\bibinfo{author}{B.~A. Volkov}, \bibinfo{author}{O.~A. Pankratov},
  \bibinfo{journal}{Sov. Phys. JETP} \bibinfo{volume}{42}
  (\bibinfo{year}{1985}) \bibinfo{pages}{178}.
\bibitem[{Korenman and Drew(1987)}]{Korenman87}
\bibinfo{author}{V.~Korenman}, \bibinfo{author}{H.~D. Drew},
  \bibinfo{journal}{Phys. Rev. B} \bibinfo{volume}{35} (\bibinfo{year}{1987})
  \bibinfo{pages}{6446}.
\bibitem[{Agassi and Korenman(1988)}]{Agassi88}
\bibinfo{author}{D.~Agassi}, \bibinfo{author}{V.~Korenman},
  \bibinfo{journal}{Phys. Rev. B} \bibinfo{volume}{37} (\bibinfo{year}{1988})
  \bibinfo{pages}{10095}.
\bibitem[{Pankratov(1990)}]{Pankratov90}
\bibinfo{author}{O.~A. Pankratov}, \bibinfo{journal}{Semicond. Sci. Technol.}
  \bibinfo{volume}{5} (\bibinfo{year}{1990}) \bibinfo{pages}{S204}.
\bibitem[{Kolesnikov and Silin(1997)}]{Kolesnikov97}
\bibinfo{author}{A.~V. Kolesnikov}, \bibinfo{author}{A.~P. Silin},
  \bibinfo{journal}{J. Phys.: Condens. Mat.} \bibinfo{volume}{9}
  (\bibinfo{year}{1997}) \bibinfo{pages}{10929}.
\bibitem[{Kriechbaum(1986)}]{Kriechbaum86}
\bibinfo{author}{M.~Kriechbaum}, \bibinfo{title}{Envelope Function Calculations
  for Superlattices}, \bibinfo{publisher}{Springer}, \bibinfo{address}{Berlin},
  \bibinfo{year}{1986}, p. \bibinfo{pages}{120}.
\bibitem[{Dom{\'i}nguez-Adame(1994)}]{Adame94}
\bibinfo{author}{F.~Dom{\'i}nguez-Adame}, \bibinfo{journal}{phys. stat. sol.
  (b)} \bibinfo{volume}{186} (\bibinfo{year}{1994}) \bibinfo{pages}{K49}.
\bibitem[{Hwang et~al.(2012)Hwang, Siegel, Mo, Regan, Ismach, Zhang, Zettl, and
  Lanzara}]{Hwang12}
\bibinfo{author}{C.~Hwang}, \bibinfo{author}{D.~A. Siegel},
  \bibinfo{author}{S.-K. Mo}, \bibinfo{author}{W.~Regan},
  \bibinfo{author}{A.~Ismach}, \bibinfo{author}{Y.~Zhang},
  \bibinfo{author}{A.~Zettl}, \bibinfo{author}{A.~Lanzara},
  \bibinfo{journal}{Sci. Rep.} \bibinfo{volume}{2} (\bibinfo{year}{2012})
  \bibinfo{pages}{590}.
\bibitem[{Li et~al.(2010)Li, Luican, Lopes~dos Santos, Castro~Neto, Reina,
  Kong, and Andrei}]{Li10}
\bibinfo{author}{G.~Li}, \bibinfo{author}{A.~Luican}, \bibinfo{author}{J.~M.~B.
  Lopes~dos Santos}, \bibinfo{author}{A.~H. Castro~Neto},
  \bibinfo{author}{A.~Reina}, \bibinfo{author}{J.~Kong}, \bibinfo{author}{E.~Y.
  Andrei}, \bibinfo{journal}{Nat. Phys.} \bibinfo{volume}{6}
  (\bibinfo{year}{2010}) \bibinfo{pages}{109}.
\bibitem[{Trambly~de Laissardi\`ere et~al.(2010)Trambly~de Laissardi\`ere,
  Mayou, and Magaud}]{Trambly10}
\bibinfo{author}{G.~Trambly~de Laissardi\`ere}, \bibinfo{author}{D.~Mayou},
  \bibinfo{author}{L.~Magaud}, \bibinfo{journal}{Nano Letters}
  \bibinfo{volume}{10} (\bibinfo{year}{2010}) \bibinfo{pages}{804}.
\bibitem[{Das et~al.(2015)Das, Sarkar~Manna, and Mitra}]{Das15}
\bibinfo{author}{D.~Das}, \bibinfo{author}{J.~Sarkar~Manna},
  \bibinfo{author}{M.~K. Mitra}, \bibinfo{journal}{J. Phys. Chem. C}
  \bibinfo{volume}{119} (\bibinfo{year}{2015}) \bibinfo{pages}{6939}.
\bibitem[{Elias et~al.(2011)Elias, Gorbachev, Mayorov, Morozov, Zhukov, Blake,
  Ponomarenko, Grigorieva, Novoselov, Guinea, and Geim}]{Elias11}
\bibinfo{author}{D.~C. Elias}, \bibinfo{author}{R.~V. Gorbachev},
  \bibinfo{author}{A.~S. Mayorov}, \bibinfo{author}{S.~V. Morozov},
  \bibinfo{author}{A.~A. Zhukov}, \bibinfo{author}{P.~Blake},
  \bibinfo{author}{L.~A. Ponomarenko}, \bibinfo{author}{I.~V. Grigorieva},
  \bibinfo{author}{K.~S. Novoselov}, \bibinfo{author}{F.~Guinea},
  \bibinfo{author}{A.~K. Geim}, \bibinfo{journal}{Nat. Phys.}
  \bibinfo{volume}{7} (\bibinfo{year}{2011}) \bibinfo{pages}{701}.
\bibitem[{Miao et~al.(2013)Miao, Wang, Ming, Yao, Wang, Yang, Song, Zhu,
  Fedorov, Sun, Gao, Liu, Xue, Liu, Liu, Qian, and Jia}]{Miao13}
\bibinfo{author}{L.~Miao}, \bibinfo{author}{Z.~F. Wang},
  \bibinfo{author}{W.~Ming}, \bibinfo{author}{M.-Y. Yao},
  \bibinfo{author}{M.~Wang}, \bibinfo{author}{F.~Yang}, \bibinfo{author}{Y.~R.
  Song}, \bibinfo{author}{F.~Zhu}, \bibinfo{author}{A.~V. Fedorov},
  \bibinfo{author}{Z.~Sun}, \bibinfo{author}{C.~L. Gao},
  \bibinfo{author}{C.~Liu}, \bibinfo{author}{Q.-K. Xue}, \bibinfo{author}{C.-X.
  Liu}, \bibinfo{author}{F.~Liu}, \bibinfo{author}{D.~Qian},
  \bibinfo{author}{J.-F. Jia}, \bibinfo{journal}{Proc. Natl. Acad. Sci.}
  \bibinfo{volume}{110} (\bibinfo{year}{2013}) \bibinfo{pages}{2758}.
\bibitem[{D{\'i}az-Fern{\'a}ndez et~al.(2017)D{\'i}az-Fern{\'a}ndez, Chico,
  Gonz{\'a}lez, and Dom{\'i}nguez-Adame}]{Diaz-Fernandez17}
\bibinfo{author}{A.~D{\'i}az-Fern{\'a}ndez}, \bibinfo{author}{L.~Chico},
  \bibinfo{author}{J.~W. Gonz{\'a}lez},
  \bibinfo{author}{F.~Dom{\'i}nguez-Adame}, \bibinfo{journal}{ArXiv e-prints}
  (\bibinfo{year}{2017}). \href{http://arxiv.org/abs/1702.08296}{\tt
  arXiv:1702.08296\normalfont}.
\bibitem[{Alberto et~al.(2011)Alberto, Das, and C.~Vagenas}]{Alberto11}
\bibinfo{author}{P.~Alberto}, \bibinfo{author}{S.~Das},
  \bibinfo{author}{E.~C.~Vagenas}, \bibinfo{journal}{Phys. Lett. A}
  \bibinfo{volume}{375} (\bibinfo{year}{2011}) \bibinfo{pages}{1436}.
\bibitem[{Sauter(1931)}]{Sauter31}
\bibinfo{author}{F.~Sauter}, \bibinfo{journal}{Z. Physik} \bibinfo{volume}{69}
  (\bibinfo{year}{1931}) \bibinfo{pages}{742}.
\bibitem[{Abramowitz and Stegun(1972)}]{Abramowitz72}
\bibinfo{author}{M.~Abramowitz}, \bibinfo{author}{I.~Stegun},
  \bibinfo{title}{Handbook of Mathematical Functions},
  \bibinfo{publisher}{Dover}, \bibinfo{address}{New York},
  \bibinfo{year}{1972}.

\end{thebibliography}

\bibliographystyle{model1a-num-names.bst}

\end{document}